# Intrinsically Legal-For-Trade Objects by Digital Signatures

A. Wiesmaier[†]  U. Rauchschwalbe[∗]  C. Ludwig[†]  B. Henhapl[†]  M. Ruppert[‡]  J. Buchmann[†]

| [†]TU Darmstadt – CDC | [∗]Schenck Process GmbH | [‡]FlexSecure GmbH |
|---|---|---|
| Hochschulstr. 10 | Landwehrstr. 55 | Industriestr. 12 |
| D–64289 Darmstadt | D–64293 Darmstadt | D–64297 Darmstadt |

**Abstract:** The established techniques for legal-for-trade registration of weight values meet the legal requirements, but in praxis they show serious disadvantages. We report on the first implementation of intrinsically legal-for-trade objects, namely weight values signed by the scale, that is accepted by the approval authority. The strict requirements from both the approval- and the verification-authority as well as the limitations due to the hardware of the scale were a special challenge. The presented solution fulfills all legal requirements and eliminates the existing practical disadvantages.

## 1 Introduction

Scales used in business transactions must have a type approval. They are verified and are subject to periodic supervision. The German weights and measures act demands – beside the reliability of the scale – that the customer is able to check that the weighing was accurate. In many business processes the customer cannot attend the actual weighing process which normally leads to the demand for a *legal-for-trade (lft) registry*. A registration can be lft if the storage of the weight values is triggered and executed by a device that is subject to legal control.

Traditionally, lft registration is realized by *alibi printers*, i. e., devices that are directly attached to the scales and generate hardcopies of all executed weighings. However, alibi printers are unpopular among the scale operators, since they are noisy, need maintenance, and cost space as well as money. Scales with *alibi memory* were introduced to remedy these drawbacks: Such scales keep the weight values for a sufficiently long time in some kind of internal memory (e. g., EEPROM). (Usually a storage time of 90 days is considered sufficient.) Alibi storage comes with its own set of problems, though: Since more memory makes the scales more expensive, the memory of electronic scales is typically scarce. The displays and keyboards attached to the scales are often very small, whence reading the stored values is uncomfortable. Therefore, the next generation of lft registries relocated the alibi memory into computer systems connected to the scale. Again, the elimination of existing problems generated new ones. The involved software on the computer systems needs official approval and is subject to supervision by the authorities. Securing the software is relatively complex and the solution is bound to a specific operating system — in most cases Windows.

All described solutions have the handicap in common that the customer can check the data only at the site where scale is installed. This explains why customers take their chance for verification very rarely. On the other hand, any data that is transmitted to the customer leaves the verified environment. Facing the increasing global trade relations, this is a substantial disadvantage. This situation motivated us to search for a new method that is platform independent, does not cause operating costs, and allows weight data verification at the consignee's site.

The basic idea is as follows: Instead of filing the measured weights in an approved device that stores and displays the data, we turn the weight values themselves into intrinsically lft objects by means of a digital signature. The lft registry therefore becomes redundant. After its generation in the scale, the signed record can be saved or transported using arbitrary media not subject to legal control. In principle, verification of the data is possible at any site.

## 2  Situation

Some circumstances were favourable for the project. Modern, efficient algorithms for digital signatures are known and freely available, in particular free of license fees. The European Union directive 1999/93/EC [Eur99], in Germany implemented by the German signature act [Sig01], provides an international legal base for the utilization of such procedures. The Institute for Cryptography and Computer Algebra at the TU Darmstadt and FlexSecure have comprehensive knowledge and broad experience in this field. Schenck Process complements this with its experiences in the field of weighing technology and the related laws. The Physikalisch-Technische Bundesanstalt (PTB) – the approving authority – has experiences from its own similar project called SELMA that deals with gas and water meters. The PTB signaled its willingness to approve this technology for the use in scales.

The technical constraints were less favorable. The available scale hardware works with a 16 bit processor (Infineon SAB C167) at a frequency of 40 MHz and offers no cryptographic co-processor. So the CPU is several orders of magnitude less powerful than CPUs of commercial PCs. The scale possesses 1 MByte Flash-EEPROM for program code and 265 KByte RAM for data, but only 64 KByte Flash-EEPROM and 16 KByte RAM are available for the digital signatures. The rest is needed for the operating system and the other features of the scale. Hard discs or other drives are not available. Nevertheless, a signature operation must always complete within 10 seconds. This period corresponds to the smallest interval between two weightings with scales of that type.

The concept of securing lft data by digital signatures breaks new regulatory ground. There exist neither national nor international regulations in this field. Thus, it was no surprise that the PTB and the verification authorities had some technical reservations. Intensive discussions finally led to solutions that were acceptable for all parties.

## 3 Technical Details

The scale's processor is within the verified area, i. e. direct access to the CPU is not possible without breaching the authority's seal; Furthermore, the scale runs a multi tasking system with more than 20 concurrent processes; while computing a signature, several 1000 task switches are done. Both facts hamper side channel attacks that try to recover key material, e. g., by measuring the scale's electric power consumption or radiation.

The random numbers generated by the scale are produced by repetitively sampling the least significant bit of the scale's analog digital converter. The resolution of the converter is much higher than its accuracy. This RNG complies, e. g. with FIPS 140-1.

The software generates ECDSA signatures using elliptic curves over $GF(p)$. It relies on the long integer and elliptic curve arithmetic in the MIRACL library [SSL] that performs well on our platforms. The key length is 160 bit. The curve parameters are embedded in the program code and are therefore shared by all scales. In order to avoid any patent infringements, the parameters were generated by the implementation of the complex multiplication method in the open source library LiDIA [LG04].

The key pair is generated by the scale at the producer's site and stored in the scale's EEPROM before its delivery. The private key never leaves the scale. The scale's memory is not readable from the outside without breaking the seal.

Upon startup, the device precomputes parameters for the point multiplication to speed up later signatures. Furthermore, the scale maintains a pool of precomputed ephemeral ECDSA keys. This pool is refilled by a background process.

The trust anchor is a self signed root certificate. The private root key (RSA-1024) resides on a PIN protected chip card. Running the trust center software requires that at least 2 operators – authenticated by their own PIN protected chip cards – are logged in.

## 4 Procedure

At the manufacturer's site, each scale is attached to a computer that is dedicated to running the trust center software (FlexiTRUST) only. This computer issues the command to generate a key pair and the scale responds with the public key. The trust center software creates an X.509 certificate that binds the public key to the scale's serial number and publishes the certificate in a publicly accessible LDAP directory.

The signature computation after each weighting takes about 1 second and is thus not noticed by the scale operator as an additional delay. The signature has a size of 2 x 160 bit and is converted into a hexadecimal representation of 2 x 40 characters to facilitate printing. The scale outputs the signature together with the weight data in both a clear text representation for the consignee as well as a (configurable) compressed representation. (The latter representation eases the input of the data into verification programs.) After 5 - 10 seconds (depending on the system load) the scale is ready for the next weighting. Note that we can at any time add redundant information (e. g., for error correcting) or recode the data as long as the original data and signature can be recovered for verification.

Both the signature algorithm and the used curve parameters are public knowledge. All certificates are accessible via the Internet. Thus, any customer can verify the signed data records with the tools of his choice. Several solutions suitable for the use by laymen exist:

First, the user copies the signed data record into a web form whereupon the system checks the signature and the corresponding certificate. Due to the likelihood of typos this service is suitable if the customer needs to verify few signatures only. Second, we developed a standalone program that can be run at the consignee's site. It allows the manual data input similar to the web service as well as the import of the compressed data record. The program can fetch the necessary certificates from the LDAP directory and caches them locally, so it does not need permanent Internet access. Finally, our data records are also compatible with a verification application developed by the PTB within its SELMA project. The availability of a verification tool from an independent party conduces to the acceptance of our system by the approval authority and customers.

## 5 Outlook

A temporary approval of the system is under way. This approval will allow all involved parties (manufacturer, customers, approving authority, and measurement office) to gain experience with digitally signed legal-for-trade objects on a firm legal fundament. Patents for the system are pending; the system was already disclosed within the scope of the patent registration. The system is marketed by Schenck Process.

We expect that the speed up due to the next upcoming scale hardware upgrade will lead to noticeably shorter signature computation times even if we use 192 bit keys. Given the increasing usage of PCs in the field of lft scale technology, we are interested in the transfer of our system to the open PC architecture. We are reviewing potential solutions that meet the regulatory requirements.

## References

[Eur99] European Union Parliament and Council. Directive on a Community Framework for digital Signatures; 1999/93/EC, 1999. http://www.e-podpis.sk/laws/eu_ep_dir93_1999.pdf (26 Sep. 2005).

[LG04] LiDIA-Group. LiDIA – A Library for Computational Number Theory. TU Darmstadt, 2004. http://www.informatik.tu-darmstadt.de/TI/LiDIA/Welcome.html.

[Sig01] Gesetz über Rahmenbedingungen für elektronische Signaturen und zur Änderung weiterer Vorschriften. *Bundesgesetzblatt Jahrgang 2001 Teil I*, 22:876–884, 2001. http://bundesrecht.juris.de/bundesrecht/sigg_2001/ (26 Sep. 2005).

[SSL] Shamus Software Ltd. Multiprecision Integer and Rational Arithmetic C/C++ Library. http://indigo.ie/~mscott/ (26 Sep. 2005).